\documentstyle[12pt,aaspp]{article}
\lefthead{PRESS, RYBICKI, \& SCHNEIDER}
\righthead{HIGH-REDSHIFT LYMAN ALPHA CLOUDS}

\begin{document}

\def\pslandinsert#1{\epsffile{#1}}
\def\eqref#1{\ref{eq#1}}
\def\e#1{\label{eq#1}}
\def\be{\begin{equation}}
\def\ee{\end{equation}}
\def\ast{\mathchar"2203} \mathcode`*="002A
\def\rslash{\backslash} \def\oforder{\sim}
\def\larrow{\leftarrow} \def\rarrow{\rightarrow}
\def\darrow{\Longleftrightarrow}
\def\defeq{\equiv} \def\lteq{\leq} \def\gteq{\geq} \def\neq{\not=}
\def\<={\leq} \def\>={\geq} \def\lsls{\ll} \def\grgr{\gg}
\def\all{\forall} \def\lub{\sqcup} \def\relv{\vert}
\def\leftv{\left|} \def\rightv{\right|}
\def\%{\char'045{}}
\def\_{\vrule height 0.8pt depth 0pt width 1em}
\def\leftbrace{\left\{} \def\rightbrace{\right\}}
\def\sectsign{\S}
\def\prop{\propto}
\newbox\grsign \setbox\grsign=\hbox{$>$} \newdimen\grdimen \grdimen=\ht\grsign
\newbox\simlessbox \newbox\simgreatbox
\setbox\simgreatbox=\hbox{\raise.5ex\hbox{$>$}\llap
     {\lower.5ex\hbox{$\sim$}}}\ht1=\grdimen\dp1=0pt
\setbox\simlessbox=\hbox{\raise.5ex\hbox{$<$}\llap
     {\lower.5ex\hbox{$\sim$}}}\ht2=\grdimen\dp2=0pt
\def\simgreat{\mathrel{\copy\simgreatbox}}
\def\simless{\mathrel{\copy\simlessbox}}
\def\spose{\rlap} \def\caret#1{\widehat #1}
\def\hat#1{\widehat #1} \def\tilde#1{\widetilde #1}
\def\limitswitch{\limits} \def\dispstyle{\displaystyle}
\def\dot#1{\vbox{\baselineskip=-1pt\lineskip=1pt
     \halign{\hfil ##\hfil\cr.\cr $#1$\cr}}}
\def\ddot#1{\vbox{\baselineskip=-1pt\lineskip=1pt
     \halign{\hfil##\hfil\cr..\cr $#1$\cr}}}
\def\dddot#1{\vbox{\baselineskip=-1pt\lineskip=1pt
     \halign{\hfil##\hfil\cr...\cr $#1$\cr}}}
\def\Abf{{\bf A}}\def\ybf{{\bf y}}\def\Ebf{{\bf E}}\def\cbf{{\bf c}}
\def\Zbf{{\bf Z}}\def\Bbf{{\bf B}}\def\Cbf{{\bf C}}\def\bfB{{\bf B}}
\def\bfq{{\bf q}}\def\bfc{{\bf c}}\def\bfy{{\bf y}}\def\bfv{{\bf v}}
\def\bfp{{\bf p}}

\def\Lya{Lyman $\alpha$}
\def\ang{\hbox{\AA}}
\def\calN{{\cal N}}
\def\Mbar{{\overline M}}

\title{PROPERTIES OF HIGH-REDSHIFT\\
       LYMAN ALPHA CLOUDS\\
       I. STATISTICAL ANALYSIS OF THE SSG QUASARS}

\author{William H. Press and George B. Rybicki}
\affil{Harvard-Smithsonian Center for Astrophysics,
     Cambridge, MA 02138}

\and

\author{Donald P. Schneider}
\affil{Institute for Advanced Study, Princeton, NJ 08540}


\begin{abstract}

Techniques for the statistical analysis of the \Lya\ forest in high
redshift quasars are developed, and applied to the low resolution
(25 \AA) spectra of 29 of the 33 quasars in the Schneider-Schmidt-Gunn
(SSG) sample.  We extrapolate each quasar's continuum shortward of \Lya\
emission, then consider each spectral bin of each quasar to be an
(approximately) independent measurement of the absorption due to the
\Lya\ clouds.  With several thousand such measurements thus available, we can
obtain good determinations of some interesting properties of clouds in
the redshift range $2.5 < z < 4.3$ without actually resolving any
single cloud.  We find that the mean absorption increases with $z$
approximately as a power law $(1+z)^{\gamma+1}$ with $\gamma = 2.46\pm
0.37$.  The mean ratio of \Lya\ to Lyman $\beta$ absorption in the
clouds is $0.476\pm 0.054$.  We also detect, and obtain ratios, for
Lyman $\beta$, $\gamma$, and possibly $\epsilon$. We are also able to
quantify the fluctuations of the absorption around its mean, and find
that these are comparable to, or perhaps slightly larger than, that
expected from an uncorrelated distribution of clouds. The
techniques in this paper, which include the use of bootstrap
resampling of the quasar sample to obtain estimated errors and error
covariances, and a mathematical treatment of absorption from a
(possibly non-uniform) stochastic distribution of lines, should be
applicable to future, more extensive, data sets.

\end{abstract}


\clearpage

\section{Introduction}\label{I}

Excepting only the quasars themselves, no observable objects have more
potential for revealing the quantitative nature of the
Universe at early times than do the \Lya\ clouds, seen in
absorption against the UV continuum of background quasars.  Indeed,
one might argue that the importance of the \Lya\ clouds,
at redshifts $z>2.5$, in some respects
exceeds that of the quasars, since the high-redshift quasars are
``unusual'' objects, presumably associated with the extreme statistical
tail of structure formation, while the \Lya\ clouds are probably more
``typical'' representatives of the state of baryonic matter, at least in the
redshift range $2.5 < z < 4.3$.

{}From the time of their first discovery (Lynds 1971) and first
detailed analysis (Sargent et al.~1980), the \Lya\ clouds have
principally been studied by the spectroscopic identification, and
counting, of individual clouds along a quasar line of sight.
(For a review, see, e.g., Sargent 1988a.) Since
the cloud absorption lines are narrow, with equivalent widths on
the order of 1 \ang\ or less, observations with moderate-to-high
spectral resolving power are required.  Indeed, some exceptionally
high quality QSO spectra (resolution $\simless 0.1 \ang$)
have been obtained for use in direct studies of the cloud line profiles
and velocity dispersion parameters (Pettini et al.~1990,
Carswell et al.~1991).

At the opposite extreme lies the set of techniques that study the
gross relative depression of the quasar continuum spectrum shortward
of \Lya\ (that is, emitted wavelength $\lambda_{em} < 1216 \ang$) by
the aggregate effect of many clouds along the line of sight at
redshifts smaller than the emission redshift $z_{em}$ (Oke and
Korycansky 1982; Bechtold et al.~1984; Schneider et al.  1989a,b,
1991; Giallongo and Cristiani 1990;
Jenkins and Ostriker, 1991).  Typical of these techniques is the
association of a single number $D_A$ with each quasar line of sight,
the fractional absorption averaged over a broad band safely between the
\Lya\ and Lyman $\beta$ emission wavelengths, say
$1050 \ang < \lambda_{em} < 1170 \ang$.

Despite the obvious fundamental importance of understanding the nature
of the intergalactic medium at high redshifts ($z>3.5$, say), there
have been very few high-redshift
studies (Schneider et al. 1989b, 1991; Jenkins
and Ostriker, 1991), and these have all used the $D_A$ approach --
because the quasars available for study are so faint.

In this paper we develop and apply a new technique, intermediate
between the above extremes (though in most ways closer to the latter).
Our data set consists of the carefully calibrated,
low resolution spectra of 33 high-redshift quasars obtained by
Schneider, Schmidt, and Gunn (1991; hereafter ``SSG'').
This sample provides the best available data set for $z>3.5$ because
(i) it contains all of the quasars with published redshifts $z>3.85$
except PC 1247+3406 ($z=4.897$, too large for this study), (ii)
the spectra were all obtained with the same instrumentation and thus
have similar resolution and noise properties, and (iii) the sample
contains some redshifts below $z=3.5$, so there is some overlap with
``detailed'' studies.  Our new approach (but cf. Webb et al. 1992) is
to consider each separately resolved wavelength of each individual quasar
to be an independent, albeit noisy, measurement of the cloud absorption
in a particular three-dimensional volume of the Universe.
(For $\lambda_{em} < 1025\ang$, the emission wavelength of Lyman $\beta$,
each volume consists of more than one disjoint piece.)

We will find that, for the 25\ang\ spectral resolution of the SSG
data, there are typically several or dozens, but not hundreds, of
significantly absorbing
clouds in each resolution element.  Since, however, we have not 29
measured numbers, but rather several thousand, it is possible to
untangle the statistics of the overlapping clouds to quite a
remarkable degree.  We will see, for example, that it is possible to
obtain meaningful measurements of not only \Lya\ absorption, but also
Ly-$\beta$ (1025\ang), Ly-$\gamma$ (972\ang), Ly-$\delta$ (949\ang), and
possibly Ly-$\epsilon$ (937\ang).

Jenkins and Ostriker (1991) showed that useful information could
potentially be obtained from the detailed distribution
of absorption values seen (at low or moderate resolution) in
the \Lya\ forest of a single quasar's line of sight.  We extend
that important idea in this paper, and show that, from
statistics of the forest's fluctuating absorption, one can derive
fairly precise measurements of several statistical quantities
associated with the \Lya\ cloud distribution.

The output of this paper is a set of techniques for measuring several
quantitative statistical properties of the \Lya\ cloud
distribution, along with detailed uncertainty estimates (including
cross-correlations), and also the results obtained by applying
these techniques to the SSG sample quasars at redshifts $2.6 < z_{abs}
< 4.2$.  Paper II of this series will show that these
measured values are already precise enough to impose strong
constraints on the physical nature of individual clouds and on the
distribution function of the cloud population.  Using these derived
observational constraints, Paper III will be able to confront the
grander cosmological questions associated with the clouds: their
origin, confinement mechanism, and implications for the formation of
galaxies and large scale structure.

The focus of this paper, and later papers in this series, is on those
\Lya\ clouds at sufficiently high redshift to be considered (at least
potentially) primordial cosmological objects, not associated with
galaxies and minimally polluted by stellar element production.  It is
already clear (see, e.g, Sargent 1988b, Bahcall et al. 1992) that the
detailed properties of quasar absorption clouds at low redshifts may
differ significantly from those at high redshifts.  We do not expect
the results of this paper to apply at low redshifts; the question
``how low is low?'', obviously an interesting one, will be considered
in a later paper.

In this paper, \S 2 discusses how we fit the individual quasar
spectra to obtain the underlying continua.  In \S 3 we estimate
the mean optical depth in \Lya\ absorption as a function of redshift $z$.
\S 4 extends this estimate to the other accessible Lyman lines.
\S 5 analyzes the fluctuation statistics of the absorption measurements.
\S 6 summarizes our conclusions.

In Appendix A we analyze in detail a class of statistical
models for the distribution of spectral lines in the Lyman forest.
These are similar to the usual Poisson processes, but
allow for a nonuniform underlying density of points.
We obtain results for the mean of the transmission coefficient and
also for its correlation function.  In Appendix B the results of
Appendix A are specialized for the particular applications of this
paper.

\section{Estimating the Continuum}\label{II}

The input data to this study are the calibrated SSG spectra of 33
quasars in the redshift range of $3.1 < z_{em} < 4.8$.  The data are
in the form of corrected (for atmospheric absorption, reddening, and
instrumental response) fluxes, in 10 \AA\ bins, from 4310 \AA\ to 9500
\AA\ observed wavelength.  The resolution of the measurements is about
25 \AA, so neighboring bins are not independent.  (We make use of this
oversampling below.)  Details of the observations and data reduction
procedures are described in SSG.

Three of the SSG quasars are broad absorption line (BAL) quasars.  We
eliminate these from the sample.  At a later stage of our processing,
we also eliminate one additional quasar, the highest redshift member
of the sample, because its available data do not adequately determine
certain fitted parameters (see below). The 29 remaining quasars, along
with their redshifts and monochromatic $AB$ magnitudes at emitted
wavelength 1450 \AA\ as determined by SSG, are listed in Table 1.
(Emitted wavelength 1450 \AA, in the gap between the SiIV/OIV blend at
1400 \AA\ and CIV at 1549 \AA, is chosen as a relatively clean measurement
of the underlying quasar continuum.)

\begin{table}
\begin{center}
\begin{tabular}{|lll|lll|}  \hline\hline
{\em Quasar} &  $z$    &  $AB_{1450}$ &
  {\em Quasar} &  $z$    &  $AB_{1450}$ \\ \hline
PC2344+0124  &  3.143  &  19.10  &
  PC2047+0123  &  3.799  &  19.23  \\
PC0056+0125  &  3.149  &  18.41  &
  PC1643+4631B  &  3.831  &  20.35  \\
PC1601+3754  &  3.188  &  19.68  &
  PC1301+4747  &  4.004  &  21.32  \\
PC2132+0126  &  3.194  &  19.69  &
  Q0046$-$293  &  4.014  &  19.26  \\
PC1605+4631  &  3.203  &  19.79  &
  PC0910+5625  &  4.035  &  20.86  \\
PC0118+0119  &  3.241  &  19.19  &
  Q0101$-$304  &  4.072  &  19.98  \\
PC2226+0216  &  3.273  &  18.80  &
  PC2331+0216  &  4.093  &  19.84  \\
PC0234+0120  &  3.300  &  19.96  &
  Q0000$-$26  &  4.098  &  17.46  \\
PC0344+0222  &  3.377  &  20.09  &
  PC0104+0215  &  4.171  &  19.67  \\
PC1619+4631  &  3.471  &  20.54  &
  PC0751+5623  &  4.281  &  19.65  \\
PC1548+4637  &  3.544  &  19.07  &
  PC0307+0222  &  4.379  &  19.92  \\
PC0345+0130  &  3.638  &  19.49  &
  Q2203+29  &  4.399  &  20.41  \\
PC1640+4628  &  3.700  &  19.29  &
  PC1233+4752  &  4.447  &  20.11  \\
PC1643+4631A  &  3.790  &  20.05  &
  PC0953+4749  &  4.457  &  19.09  \\
PC0131+0120  &  3.792  &  19.08  &  &  &  \\ \hline\hline
\end{tabular}
\end{center}
\caption{The 29 quasars studied in this paper.  Redshifts and
corrected continuum $AB$ magnitudes at emitted wavelength
1450 \AA\ are from Schneider, Schmidt, and Gunn (1991).  Four SSG quasars,
the three broad
absorption line quasars, and quasar PC1158+4635 at redshift $z=4.733$,
are omitted from this study (see text).}
\end{table}

While quasar studies frequently focus on differences among quasar
spectra, for our purposes it is equally important to take note of the
similarity of all the spectra in this sample.  Figure 1 plots
all of the SSG measurements of all 29 quasars, as a function
of emitted wavelength in the range 930 \AA\ to 2200 \AA,
normalized to an (arbitrary) common $AB_{1450}$ magnitude of $18$.
One sees clearly all of the spectral features present in low redshift
quasars (see, e.g., Francis, et al. 1991).  The \Lya\ depression
shortward of 1216 \AA\ is seen with equal clarity.

While the logarithmic slope of the underlying continuum indisputably
varies from quasar to quasar, it is striking, in Figure 1, that this
is a relatively small effect over the wavelength range shown.  The
reliability of our results will depend somewhat on the accuracy with
which we are able to extrapolate each quasar's underlying continuum
down to $\sim 930$\AA.  Figure 1 shows that this extrapolation,
amounting to less than 25\% of the range of abscissa shown, is not so
daunting as one might think.

To fit for the continua, we use all the available measurements between
$\lambda_{em} = 1250$\AA\ and $\lambda_{em} = 2200$\AA.  For each
quasar in the sample, we proceed as follows.  First, we estimate a
relative statistical error associated with each measured data point.
This is done by (a) calculating the variance of each point with its
immediately adjacent neighbors, and (b) convolving this series of raw
variances with a Gaussian profile of width about 12 bins (FWHM).
(Here is where we exploit the fact that the data is oversampled.)  It
is not important that this estimate of statistical error be correctly
normalized, but only that it reflect, in a general way, the relative
weight to be assigned to individual measurements in the next stage
of fitting.

Now, we fit the data to the following 22-parameter {\em linear} model
in the emitted rest frame,
\be
F(\lambda) = C_{1/2} \lambda^{1/2} + C_1 \lambda +
  \sum_{i=1}^{10} \left[ A_i + B_i
  \left({\lambda-\lambda_i\over 2 w_i}\right)^2
  \right] \exp\left[ - \left({\lambda-\lambda_i\over 2 w_i}\right)^2 \right]
\e{1}\ee
for the parameters $C_{1/2}$, $C_1$, $A_1\ldots A_{10}$ and $B_1\ldots
B_{10}$.  Parameters $C_{1/2}$ and $C_1$ characterize the underlying
continuum with two degrees of freedom; over our limited wavelength
range they contain information equivalent to a magnitude and a spectral index.
The constant values $\lambda_i$ and $w_i$, $i=1,\ldots,10$, are the
wavelengths and {\it nominal} widths of the 10 fitted lines, and are
given in Table 2.  Parameters $A_1\ldots A_{10}$ are the fitted strengths of
the emission lines, while $B_1\ldots B_{10}$ parametrize
the ratio of the fitted width to the
nominal width.  The nominal widths in Table 2 were obtained by
fitting a nonlinear model to the composite spectrum of Figure 1.

\begin{table}
\begin{center}
\begin{tabular}{|lll|}  \hline\hline
{\em Line} & $\lambda_0$ (\AA) & $w_0$ (\AA) \\ \hline
Ly$\alpha$ & 1216. & 20. \\
NV & 1240. & 18. \\
OI & 1302. & 7.2 \\
CII & 1335. & 6.0 \\
SiIV/OIV & 1400. & 11. \\
CIV & 1549. & 15. \\
HeII/OIII & 1651. & 77. \\
FeII feature & 1770. & 26. \\
AlIII & 1858. & 30. \\
CIII & 1909. & 22. \\  \hline\hline
\end{tabular}
\end{center}
\caption{Lines, emission wavelengths, and nominal widths used for
fitting the underlying continua of individual quasars in the
range 1250 \AA\ to 2200 \AA. Nominal widths are used only as the
starting point for a linearized width correction.}
\end{table}

The functional form of equation (\eqref{1}) is perhaps slightly
unconventional, and is motivated by the desire that it be linear in
all parameters.  More conventionally, one might fit nonlinearly for a
continuum magnitude and spectral index and for the intensities and
widths of Gaussian-profile lines.  However, with the present noisy
data, and the virtual certainty that there are other features in the
data besides those modeled (especially longward of 1700 \AA), we found
such nonlinear fits to be quite fussy and to require considerable user
intervention.  Since we are not here interested in details of the
shapes of lines in their wings, but we {\em are} interested in treating
the entire data set in as statistically homogeneous manner as possible,
the relative robustness of a purely linear fit is desirable.

It is at this stage that we eliminate quasar PC1158+4635 at
redshift $z=4.733$ from the sample: Because of the large redshift, its
longer wavelength emission features are not well measured, and the
above fit is ill determined.  While we could easily fit its continuum
with a smaller number of parameters, we instead choose to preserve the
homogeneity of processing procedure by eliminating the quasar entirely.

Figure 2 shows all the data in the rest frame range 1250 \AA\ to 1800
\AA, and the fitted model spectra, individually for the 29 quasars.
Also plotted for each quasar in the figure is the adopted underlying
continuum.  In many cases this is simply the function
\be
F(\lambda) = C_{1/2} \lambda^{1/2} + C_1 \lambda
\e{2}\ee
where the $C$'s are fitted parameters.  In other cases, however,
while the full model of equation (\eqref{1}) is quite well determined,
its dissection into equation (\eqref{2}) plus a remainder is quite
degenerate numerically, as evidenced by a nearly degenerate family
of fits that trade off variations in $ C_{1/2}$ and $C_1$ against
unphysical (e.g., negative) values for the line strengths.  In these
cases we refit (essentially by eye, although the process
could be automated) a continuum of the form of equation (\eqref{2}) to
the output fitted curve of (\eqref{1}), requiring the continuum to lie below
the model at selected wavelengths between emission lines.  In virtually
all cases this refitting is quite well determined, suggesting that
one could readily replace our procedure by a single linear fit with positivity
constraints; we have not, however, implemented this.

The relation between equation (\eqref{2}) and the index $\alpha$ of
the more familiar power-law continuum model,
\be
f_\nu \propto \nu^\alpha \e{81}
\ee
at some fiducial wavelength $\lambda_0$ is
\be
\alpha = - {0.5 C_{1/2}  + C_1 \lambda_0^{1/2} \over
            C_{1/2}  + C_1 \lambda_0^{1/2}} \e{82}
\ee
We find a good correlation (e.g., at $\lambda_0=1450$\AA) between
the spectral indices thus obtained and those reported in SSG; however,
our $\alpha$ values are systematically larger (smaller in magnitude) by
a few tenths, with a mean in the range $-0.5$ to $-0.6$ reported
by Richstone and Schmidt (1980), Steidel and Sargent (1987),
Warren et al. (1991), and others.

Adopting these continuum models, and extrapolating them from
1250 \AA\ down to 930 \AA, we obtain the results shown, individually by quasar,
in Figure 3.  That figure shows all SSG measurements shortward of
emitted wavelength 1200 \AA\ in relation to their respective
extrapolated continua.  All of the rest of this paper is derived from
these data.  No further use is made of the data longward of 1200 \AA.

\section{Mean Optical Depth as a Function of Redshift}\label{III}

The emitted wavelength range 1050 \AA\ to 1170 \AA\ is substantially
uncontaminated by either \Lya\ or Lyman $\beta$ emission, and also
(since it is longward of Lyman $\beta$) uncontaminated by Lyman
$\beta$ absorption by intervening clouds.  While this range could in
principle be contaminated by C IV and Mg II absorption (see, e.g.,
Meyer and York 1987), or by quasar emission lines in this wavelength
region (but see below for evidence against this), we will adopt the
conventional assumption that these effects are likely negligible at
high redshifts; probably one is seeing a virtually pure sample
of \Lya\ absorption by clouds.  The data in this range from the 29 SSG
quasars are plotted in Figure 4.  The ordinate is the ratio of
observed flux to extrapolated continuum, i.e., the fractional
transmitted flux, or transmission. The abscissa of the left panel is
emitted wavelength.  One sees a widely scattered distribution of
points.  The abscissa of the right panel is observed wavelength or
(equivalently, see top scale) absorption redshift.  Here one sees a
strong trend with redshift.  Note that the right panel plots only data
points which satisfy the wavelength cuts of the left panel, not the
vastly larger number of individual SSG measurements with observed
wavelength in the range shown (4300 \AA\ to 6500 \AA).

There are in fact a few data points (not shown) with transmission
greater than 1, i.e., with observed fluxes above the extrapolated
continuum, and a few points less than zero, i.e., where SSG's
background subtraction gives negative results.  These are obviously
defective values; to eliminate some less obvious, but likely
defective, values, we adopt slightly tighter cuts on the data, and
accept points with transmission between $0.1$ and $0.9$.  (We have
verified that our results are not sensitive to the values of these
cutoffs.) There are 1596 surviving points in the sample.

Converting transmission to an equivalent optical depth (by taking the
negative logarithm), we fit the data shown in Figure 4 a model of the
conventional form
\be
\overline{\tau}_\alpha(z) = A (1+z)^{\gamma+1}
\e{3}\ee
and obtain the best-fit values $A=0.0037$ and $\gamma = 2.46$.  The
reason for defining the exponent to be $\gamma +1$ is so that our
$\gamma$ is directly comparable to its conventional usage as the
exponent in the number distribution of clouds (see equations 2.1--2.2
of Jenkins and Ostriker 1991).

\subsection{The Bootstrap Resampling Method}

As a digression,
we here need to discuss in some detail how we obtain error estimates (or
variances) on the quantities $A$ and $\gamma$, since we will follow a
similar paradigm in obtaining variances and covariances for various
further quantities in later sections.  There are various reasons why
the formal errors that come out of the fitting procedure are
meaningless: We don't have good error estimates on the individual
measurements.  (The error estimates used in fitting the continuum were
relative, not absolute.) The individual measurements shown in Figure 4
are not statistically independent, both because there is more than one
measurement per resolution bin (adjacent measurements sample overlapping
cloud populations), {\em and} because the points associated with a
single quasar share a common source of error in the determination of
that quasar's extrapolated continuum.

A seemingly simple, yet very powerful, method for estimating the errors
is by the statistical bootstrap (or resampling) method (Efron 1982,
Efron and Tibshirani 1986, Press et al. 1992).  From the sample of
29 quasars, we generated repeated {\em resampled} sets of 29 quasars
by drawing randomly {\em with replacement}.  The typical set will thus
have on the order of 12 duplicated quasars.  No matter; we repeat
exactly the data reduction which produced the original determinations
of $A$ and $\gamma$.  The population of resulting values for $A$ and $\gamma$
can be shown (with certain technical assumptions which need not
concern us) to be distributed around the original determinations
in the same way, both variance and covariance, that the original
determinations ought to be distributed with respect to the true value.

More generally, and to be applied below, suppose
that we have some statistical procedure that determines the $M$
quantities $R_i$, $i=1,\ldots,M$, and suppose that we make $N$
resamplings.  We
denote the $k$th determination of the $i$th quantity by $R_i^k$,
$k=0,\ldots,N$, where $k=0$ uses the full sample and
$k=1,\ldots,N$ are the resamplings.  Then the best estimates for the $R_i$'s
are given by $R_i^0$, $i=1,\ldots,M$.  The standard errors $\sigma(R_i)$
are estimated by
\def\Rbar{\overline{R}}
\be
\sigma^2(R_i) \approx {1\over N}\sum_{k=1}^{N} (R_i^k-\Rbar_i)^2
\e{9}\ee
where
\be \Rbar_i \defeq {1\over N}\sum_{k=1}^{N} R_i^k
\e{10}\ee
The correlation matrix $C_{ij}$ among the $R_i^0$'s is estimated by
\be
C_{ij} \approx {1\over N}\sum_{k=1}^{N} (R_i^k-\Rbar_i)(R_j^k-\Rbar_j)
\e{11}\ee
An important quantity is the inverse correlation matrix $C_{ij}^{-1}$,
since a set of model predictions $R_i^*$ for the quantities can be compared
to the measured values $R_i^0$ by calculating
\be
\chi^2 = \sum_{i,j=1}^M (R_i^*-R_i^0) C_{ij}^{-1} (R_j^*-R_j^0)
\e{12}\ee
which should be chi-square distributed with $M$ degrees of freedom
(see, e.g., Rybicki and Press 1992).
Another sometimes useful set of quantities are the coefficients of
correlation among the individual measurements, given by
$r_{ij} \defeq C_{ij}/(\sigma_i\sigma_j)$.

We can now return to our particular example:
To obtain error bars on our measurement of $\gamma$ and $A$,
we performed 100 resamplings, obtaining the results
\be
\gamma = 2.46 \pm 0.37 \qquad A = 0.0037 \pm 0.0024
\e{4}\ee
where the errors are 1-$\sigma$.  The parameters $A$ and $\gamma$
turn out to be
very highly correlated.  If we consider $\gamma$ as the more fundamental
quantity, then almost all of the error in $A$ can be moved to the
determination of $\gamma$, as
\be
A = 0.0175 - 0.0056\gamma \pm 0.0002
\e{5}\ee
The solid line in Figure 4 shows the mean absorption of equations
(\eqref{3})--(\eqref{4}), transformed back to an ordinate of
transmission.  The shaded band shows the result of
changing $\gamma$ by $\pm 1\sigma$, with $A$ simultaneously changed
by equation (\eqref{5}).

The values of $A$ and $\gamma$ that we obtain are consistent with the
results of Jenkins and Ostriker (1991), who measure an aggregate
continuum depression $D_A$ for each quasar sampled, and in good
agreement with previous determinations which relied on the counting of
individual lines.  For example, Murdoch et al. (1986) obtained $\gamma
= 2.31\pm 0.40$; Bajtlik et al. (1988) obtained $2.36 \pm 0.40$; Lu,
Wolfe, and Turnshek (1991) obtained $2.37\pm 0.26$.  What is new in
this investigation is the extension to higher redshifts (Table 1), and
the use of a large quantity of low resolution data without line
counting and without aggregation into broad spectral bands.  Note that
the large values of $\gamma$ at high redshift are quite different from
those observed at low redshift (e.g., Bahcall et al. 1993). In \S 5,
below, and in Paper II, we will see the additional information that
comes from an absolute determination of the constant $A$, and from our
ability (not present in aggregate determinations of $D_A$) to look at
the distribution of individual points around the fitted mean values.

\section{Relative Absorption Strengths of Lyman $\beta$, $\gamma$,
   $\delta$, and $\epsilon$}\label{IV}

In Figure 5 we replot all the data shortward of $\lambda_{em} =
1200$\AA, but now removing (individually for each datum at its own
appropriate absorption redshift) the effect of the mean \Lya\
absorption, equations (\eqref{3}) and (\eqref{4}).  The ordinate is
the same as Figure 4, namely ratio of observed flux to the
extrapolated continuum.  The shaded band in the figure is simply a
moving-window average; its fluctuation scale is simply an artifact of
the chosen window size.

One sees that, shortward of the remaining tail of \Lya\ emission and
down to 1025 \AA, the corrected fluxes indeed scatter around flux
ratio unity.  There is clear O VI emission around 1032\AA, with perhaps
some contribution from Lyman $\beta$, then a visible Lyman $\beta$
decrement extending down to Lyman $\gamma$ at 972 \AA.  The $\gamma$
and $\delta$ decrements are also visible, although we will need more
definitive statistics (below) to quantify their significance.  The
decrement below Lyman $\epsilon$ at 937 \AA\, while present in the
data, will turn out to be of questionable statistical significance in
this sample.

We can now quantify the observed relative absorptions of
Lyman $\beta$, $\gamma$, $\delta$, and perhaps $\epsilon$, relative
to \Lya; in other words the ratios of the mean equivalent widths
(the precise definition of which is equation \eqref{14} below) of
different Lyman absorption lines in the clouds: $W_\beta/W_\alpha$,
$W_\gamma/W_\alpha$, $W_\delta/W_\alpha$, etc.
In the limited redshift range that we are studying,
we will assume that the equivalent width ratios do not depend on
redshift. These ratios are,
as we will see in Paper II, direct indicators of the physical
state of the clouds.

An important point is that we are {\em not} simply quantifying the
decrements seen in Figure 5.  That figure does not display the fact
that each datum has an individual emission redshift $z_{em}$
associated with it.  Use of these individual redshifts (through
equation \eqref{3}) allows, at least in principal, the removal of that
part of the scatter in Figure 5 that is due to the range of redshifts
in the quasar sample.  For example, $W_\beta/W_\alpha$ is estimated
using all points with $\lambda_{em}$ between 972 \AA\ and 1020 \AA.
Each observed point generates a one-point estimate of the form
\be
{W_\beta/\lambda_\beta \over
 W_\alpha/\lambda_\alpha}  \approx { \ln(Q^{-1})
      - {\overline{\tau}}_\alpha(\lambda_{obs}/1216\AA-1)
     \over {\overline{\tau}}_\alpha(\lambda_{obs}/1025\AA-1)}
\e{6}\ee
where $Q$ is the ratio of observed flux to the extrapolated
continuum of the point (i.e., the observed transmission),
and ${\overline{\tau}} = {\overline{\tau}}(z_{abs})$ is given
by equation (\eqref{3}). Here, $\lambda_\alpha = 1216$ \AA\ and
$\lambda_\beta = 1025$ \AA\ are the laboratory wavelengths.
The one-point estimates are then averaged.

To estimate  $W_\gamma/W_\alpha$, we use points in the range
$\lambda_{em}$ between 949 \AA\ and 972 \AA.
Now we must subtract from the measured total optical depth
$\ln(Q^{-1})$ both \Lya\ and Lyman $\beta$ corrections at their
respective absorption redshifts and relative strengths,
\be
{W_\gamma/\lambda_\gamma \over
 W_\alpha/\lambda_\alpha} \approx { \ln(Q^{-1})
      - {\overline{\tau}}_\alpha(\lambda_{obs}/1216\AA-1)
     - {\overline{\tau}}_\beta(\lambda_{obs}/1025\AA-1)
     \over {\overline{\tau}}_\alpha(\lambda_{obs}/972\AA-1)}
\e{7}\ee
where
\be
{\overline{\tau}}_\beta(z_{abs}) \defeq \left(
{W_\beta/\lambda_\beta \over
 W_\alpha/\lambda_\alpha} \right)
     {\overline{\tau}}_\alpha(z_{abs})
\e{8}\ee
is the derived mean optical depth in Lyman $\beta$ as a function of
redshift.

One continues in this manner to obtain the ratio $W_\delta/W_\alpha$
using points in the range  937 \AA\ to 949 \AA, and $W_\epsilon/W_\alpha$
using points in the range  930 \AA\ to 937\AA.  Obviously, looking at
Figure 5, one must at some point begin to wonder whether the values
obtained have any meaning.  Also, it is clear that, because of the
repeated subtraction of earlier ratios, errors in the determination
of later ratios will be highly correlated (actually, anticorrelated)
with the errors of earlier ratios.

Once again, the bootstrap technique of resampling with replacement
provides quantitative answers to these concerns.  As for the
previously determined parameters $A$ and $\gamma$, we have made 100
resampled determinations of the 4 ratios $W_\beta/W_\alpha$,
$W_\gamma/W_\alpha$, $W_\delta/W_\alpha$, and $W_\epsilon/W_\alpha$,
choosing a different set of 29 quasars in each resampling.

In the notation of the discussion leading to equation (\eqref{9}), let
$R_i$, $i=1,\ldots,4$, now denote the above four ratios, in the
obvious order.  Then the resulting measured quantities $R_i^0$,
$\sigma(R_i)$, $C_{ij}$, $C_{ij}^{-1}$, and $r_{ij}$ are given in
Table 3.  Although the last ratio, $W_\epsilon/W_\alpha$, is not
well determined (detected at less than two standard deviations), the
other ratios seem quite well established by the data.  The first two
ratios, $W_\beta/W_\alpha$ and $W_\gamma/W_\alpha$, are determined by
this data set with about 20\% accuracy.  We will see in Paper II that
these values are able to constrain the nature of \Lya\ clouds at
high redshifts that are not yet accessible to high resolution studies.

\begin{table}
\begin{center}
\begin{tabular}{|l|cccc|}  \hline\hline
{\em Ratio} & $\mathstrut\widehat W_\beta/\widehat W_\alpha$
            & $\widehat W_\gamma/\widehat W_\alpha$
            & $\widehat W_\delta/\widehat W_\alpha$
            & $\widehat W_\epsilon/\widehat W_\alpha$ \\
\quad$i=$ & 1 & 2 & 3 & 4 \\ \hline
{\em Measured Value} & 0.476 & 0.351 & 0.157 & 0.118 \\
{\em Standard Error (1-$\sigma$)\qquad} & 0.054 & 0.072 & 0.082 & 0.072 \\
\hline
\multicolumn{5}{|l|}{\em Correlation Matrix $C_{ij}$} \\
$\qquad j=1$ & 0.00296 & $-$0.00156 & $-$0.00005 & 0.00085 \\
$\qquad j=2$ & $-$0.00156 & 0.00532 & $-$0.00266 & $-$0.00041 \\
$\qquad j=3$ & $-$0.00005 & $-$0.00266 & 0.00668 & $-$0.00351 \\
$\qquad j=4$ & 0.00085 & $-$0.00041 & $-$0.00351 & 0.00522 \\ \hline
\multicolumn{5}{|l|}{\em Inverse Correlation Matrix $C_{ij}^{-1}$} \\
$\qquad j=1$ & 422. & 147. & 49.5 & $-$23.6 \\
$\qquad j=2$ & 147. & 355. & 224. & 155. \\
$\qquad j=3$ & 49.5 & 224. & 378. & 264. \\
$\qquad j=4$ & $-$23.6 & 155. & 264. & 385. \\ \hline
\multicolumn{5}{|l|}{\em Coefficient of Correlation $r_{ij}$} \\
$\qquad j=1$ & 1.00 & $-$0.39 & $-$0.01 & 0.22 \\
$\qquad j=2$ & $-$0.39 & 1.00 & $-$0.45 & $-$0.08 \\
$\qquad j=3$ & $-$0.01 & $-$0.45 & 1.00 & $-$0.59 \\
$\qquad j=4$ & 0.22 & $-$0.08 & $-$0.59 & 1.00 \\ \hline
\end{tabular}
\end{center}
\caption{Measured quantities associated with the ratios of mean equivalent
widths for Lyman $\alpha$, $\beta$, $\gamma$, $\delta$, and
$\epsilon$ in the SSG sample.  In the column headings $\widehat W_\alpha
\defeq W_\alpha/\lambda_\alpha$, $\widehat W_\beta
\defeq W_\beta/\lambda_\beta$, etc.}
\end{table}

\section{Fluctuations in Optical Depth around Mean Values}\label{V}

In principle there is information not only in the mean values of
\Lya\ absorption (and the higher lines $\beta$, $\gamma$, $\delta$),
but also in the statistics of the fluctuations of individual points
around the mean.  In Figure 5, for example, one wants to identify the
mechanisms that contribute
to the spread of the individual points around
the mean, defined in that Figure to be unity between 1050 \AA{}
and 1170 \AA.  Measurement noise, which tells us nothing about
the \Lya\ clouds,  is one contributing factor, as are uncertainties
in our extrapolation of the quasar continua, which are also not of
intrinsic interest.

The more interesting contribution, because it potentially does give
information about the clouds, is that of the Poisson statistics of
how many clouds are present in each spectral resolution element.

Unfortunately, as we will now see, the exploitation of this
information is not straightforward.

Let us first revisit the question of the mean transmission.
The absorption equivalent width $W$ of a cloud (in its rest frame)
depends on its physical parameters, for example column density $N$,
temperature parameter $b$, and so on.  Let $\bfp$ denote the vector of
such parameters, and let ${\cal N}(\bfp) d\bfp$ be the number of
clouds along the line of sight {\it per unit rest wavelength}
in a volume $d\bfp$ of parameter space.  Then the total number of clouds
per unit rest wavelength is given by
\be
n = \int {\cal N}(\bfp) d\bfp
\e{13}\ee
(if the integral converges), while the mean equivalent width of a cloud
is
\be
\overline{W} = {1\over n} \int W(\bfp){\cal N}(\bfp) d\bfp
\e{14}\ee
If the lines are randomly placed, e.g., without clustering, then
the mean transmission $\overline{Q}$ is given by
\be
\overline{Q} = \exp(-n\overline{W})
\e{15}\ee
if the integral for $n$ exists, and
\be
\overline{Q} = \exp(-\int W(\bfp){\cal N}(\bfp) d\bfp)
\e{16}\ee
otherwise (for example, if there are an infinite number of lines
with negligibly small $W$).
Equation (\eqref{15}) is so natural as to seem intuitively obvious,
but it is in fact quite a nontrivial result, relating the average
of a (nonlinear) exponential to the average of the exponent.  A proof
is given by Goody (1964, \S4.5);
see also the Appendices, equations (\eqref{A15}), (\eqref{B5}), and
(\eqref{B6}).

The sad fact is that there is  no such universal result for the next
moment,
\be
\hbox{Var}(Q)/\overline{Q}^2 = \left< \left({Q\over\overline{Q}}-1
     \right)^2\right> \e{17}
\ee
Rather, the variance $\hbox{Var}(Q)/\overline{Q}^2$ depends in a
nontrivial way on the joint distribution of equivalent widths (or, more
fundamentally, column density $N$) and doppler widths $b$, as well as
on the instrumental spectral resolution.

Some simple model cases are derived in Appendix B and are
illuminating.  Assume a square line profile of full width $W_0$, and
suppose that all lines have the same column density, and thus
the same equivalent width $\overline{W}$.
Let $\Delta$ be the instrumental spectral resolution.  Then an exact
result and its power series expansion are
\be
\hbox{Var}(Q)/\overline{Q}^2 = {2W_0\over\Delta} \left[ {1\over A}
   (e^A-1)-1\right] \approx {n\overline{W}^2\over\Delta} \left( 1 +
   {A\over 3} + \cdots \right)
\e{18}\ee
where
\be
   A \defeq {n\overline{W}^2\over W_0}
\e{19}\ee
See Appendix B, equations (\eqref{B21}) and (\eqref{B24}).
If the instrumental response is not square but is rather described by
a response function $w(\lambda-\lambda_0)$, then the definition of
$\Delta$ is
\be
\Delta = \left.  \left(\int w d\lambda \right)^2  \right/ \int w^2 d\lambda
\e{20}\ee
(cf. equation \eqref{B10.1} in Appendix B).

The limiting form of equation (\eqref{18}) for small $A$ can be
reproduced, up to a constant, by a simple physical argument: Because
the lines have a width $W_0$, the instrumental width $\Delta$ contains
$\Delta/W_0$ independent elements.  In each of these, the mean optical
depth is $n\overline{W}$, while the number of lines contributing to
this mean is $nW_0$.  Thus, in each element, the r.m.s. fluctuation in
optical depth, which is also the fractional fluctuation in
transmission $Q$ is about $(n\overline{W}) (nW_0)^{-1/2}$.  Averaging
over the independent elements reduces this fractional fluctuation by
an additional factor $(W_0/\Delta)^{1/2}$, yielding finally
\be
\hbox{Var}(Q)/\overline{Q}^2 \sim {n\overline{W}^2\over\Delta}
\e{21}\ee
The reason that the exact result (\eqref{18}) has an additional
exponentially increasing factor in $A$ is that, for large optical
depths and saturated lines, the fluctuations become dominated by the
cases where increasingly rare ``windows'' happen to occur in the
random placement of the lines.

A second analytic result derived in Appendix B (cf. equations
\eqref{B21} and \eqref{B28}) is for the case where
there is no characteristic value for $\overline{W}$, but rather a
power-law distribution in $N$,  the column density.  (This is a more
realistic idealization of the actual situation for the \Lya\ clouds.)
If the number of clouds along the line of sight with $N$ between
$N$ and $N+dN$ is proportional to $N^{-\beta}dN$, then
equation (\eqref{18}) is replaced by
\be
\hbox{Var}(Q)/\overline{Q}^2 = {2W_0\over\Delta} \left[ {1\over \kappa}
   (e^\kappa-1)-1\right]
\e{22}\ee
where now
\be
\kappa \defeq (2-2^{\beta-1}) \ln{1\over\overline{Q}}
       \defeq (2-2^{\beta-1}) \overline{\tau}
\e{23}\ee
(cf.~equations \eqref{15} and \eqref{16}).
When the mean optical depth $\overline{\tau}$ is not too large,
we have approximately
\be
\hbox{Var}(Q)/\overline{Q}^2 \approx
     (2-2^{\beta-1}){W_0\over\Delta} \overline{\tau}
\e{24}\ee
A typical value for $\beta$ is $1.5$ (Sargent 1988a; we will have more
to say about this value in Paper II).

Let us now compare these results to the SSG data.  We estimate
$\hbox{Var}(Q)/\overline{Q}^2$ in the data by the following procedure,
which is designed to greatly reduce the variance due to errors in
the continuum fits:  For each measurement $Q_i$, we estimate
$Q_i-\overline{Q}$ not by subtracting the overall mean $\overline{Q}$, but
rather by subtracting a local weighted average of the $Q_i$'s for
that particular quasar and neighboring wavelength bins.
The weighted average is taken to have a triangular profile, with unit
amplitude at bin $i$ falling to zero at bins $i\pm 30$ (that is,
$\pm 300$\AA\ in observed wavelength).  With a mean redshift of about
$1+z = 4.5$, this means that we are sensitive to variations in the
extrapolated continuum only over scales of 65 \AA, which are likely
negligible.
Our measured result for the SSG sample is
\be
\hbox{Var}(Q)/\overline{Q}^2 = 0.05 \pm 0.01
\e{25}\ee
where the 1-$\sigma$ error bar is again obtained by resampling.

At a redshift $1+z = 4.5$ we have $\overline{\tau} = 0.67$ (equation
\eqref{3}; cf.~Figure 4).  Using SSG's quoted spectral resolution
$\Delta = 25$\AA, equation (\eqref{22}) implies $W_0 \approx 0.7$\AA,
or $b \sim 80$km s$^{-1}$.  This value is implausibly large by about a
factor of about 2.  At this stage of analysis there are three possible
resolutions: (1) Since we do not have an absolute calibration of the
measurement noise for each data point, we cannot say with confidence
that this excess variance is not an instrumental effect.  (2) The
excess variance might be due to a two-point correlation function $\xi$
in the clouds on velocity scales comparable to the instrumental
resolution $\Delta$, that is $\Delta c/1216 \AA \sim 6000$km s$^{-1}$.
Clustering of \Lya\ clouds has previously been detected only at much
smaller scales (e.g., Webb 1987, Crotts 1989).
(3) The factor of 2 discrepancy might be an artifact of our simplistic
analytic assumption of a square line profile; the observed fluctuations
might in fact be consistent with a Poisson random distribution of clouds.

In Paper II, further analysis will show that the third resolution is
the most likely one.  We note here that our finding fluctuations
comparable to the predicted Poisson value helps confirm an implicit
assumption that we have made throughout this paper: the overall
absorption at high redshifts is {\it not} significantly due to a
continuous Gunn-Peterson trough rather than a superposition of
individual clouds.

\section{Conclusions}\label{VI}

The principal results of this paper are a set of techniques for
the statistical analysis of the \Lya\ forest that can be used
without the necessity of resolving and identifying individual lines.
We have seen that there exist a set of well-defined parameters that
can thus be measured with useful accuracy.  Where there is overlap
with previous results (including those obtained by high resolution
studies) our methods give a reassuring agreement.  Additionally,
our methods make accessible some new parameters, such as the
equivalent width ratios of the Lyman series.

For the redshift range of about $2.5 < z_{abs} < 4.3$ that is sampled by
the lines of sight to 29 SSG quasars, we have the following specific
results:

1.  The mean \Lya\ absorption at a redshift $z$ is well approximated
by $\overline{\tau}(z) = A (1+z)^{\gamma+1}$ with $\gamma = 2.46\pm
0.37$ and $A = 0.0175 - 0.0056\gamma \pm 0.0002$, in agreement with
previous determinations at moderate redshifts from high spectral
resolution, ground-based observations (though not with the low
redshift measurements of HST).

2.  The ratio of mean Lyman $\beta$ absorption to mean \Lya\ absorption
at a fixed absorption redshift (which is diagnostic of the physical
state of the clouds at that redshift) is $0.476\pm 0.054$.  The ratios
for Lyman $\gamma$, $\delta$, and $\epsilon$ are also measurable, and
given, along with the error covariance matrix for all the ratios, in Table 3.

3.  Fluctuations around the mean absorption are comparable to,
possibly a factor of 2 larger than can be explained as, simple Poisson
fluctuations in the number of clouds along the line of sight.  The
factor of 2 discrepancy could be due to measurement noise, model
imprecision (see Paper II), or a small, positive two-point correlation
function $\xi$ on a scale of 6000 km s$^{-1}$.  The measurement of
fluctuations comparable to the Poisson value argues against a significant
Gunn-Peterson trough at high redshifts.

\acknowledgments

We have benefited from discussions with John Bahcall and John Huchra.
The referee made a number of helpful suggestions.  This work was
supported in part by the National Science Foundation (PHY-91-06678)
and by NASA (NAG5-1618).

\appendix

\section{Appendix A}

     We here derive some statistical properties
of the extinction optical depth function $\tau(\lambda)$ as a
function of wavelength.  We base our results on
a simple statistical model, closely related to random line
models.  Anticipating needs of future papers, our development
is somewhat more general than strictly required for the present applications.

     We first express $\tau(\lambda)$ as a sum of contributions
from a number of clouds along the line of sight,
\be   \tau(\lambda) = \sum_i T(\lambda | \lambda_i,\bfp_i), \e{A2.1}\ee
where $T(\lambda | \lambda_i,\bfp_i)$ is the contribution
from the $i$th cloud.
This depends on the central wavelength of the cloud, $\lambda_i$,
as well as on a
vector of parameters $\bfp_i$ for the cloud, which may include its column
density $N_i$, its velocity parameter $b_i$, and possibly other
parameters.

     It may be helpful at this point to remark that the particular
form for $T$ used in the body of the paper
is $T(\lambda | \lambda_i,\bfp_i)=N_i \alpha(\lambda-\lambda_i,b_i)$,
where $N_i$ is the column density, $b_i$ the velocity parameter, and
$\alpha(\lambda-\lambda_i,b_i)$ is the atomic absorption coefficient,
typically a Voigt function.  In this special case
$\lambda_i$ simply defines the central wavelength of
the profile, and the parameters $\bfp_i$ consist only of $N_i$ and
$b_i$.  We shall generally assume that the $\lambda$ dependence of
$T$ falls off
sufficiently rapidly away from $\lambda_i$ to assure the convergence
of certain integrals.

     It is easiest to express our statistical model as a two-step process:
First of all, we assume that the positions of the central
wavelengths $\lambda_i$ of the clouds
are Poisson random distributed with a given mean density
of lines per unit wavelength interval.
Second, given the positions $\lambda_1$, $\lambda_2$, $\ldots$, of the clouds,
the parameters $\bfp_1$, $\bfp_2$, $\ldots$ are assumed
to be independently random, that is, their joint conditional
distribution function is a product of independent factors, one for
each cloud,
\be
  P(\bfp_1,\bfp_2,\ldots|\lambda_1,\lambda_2,\ldots)
 = \prod_i P(\bfp_i|\lambda_i).  \e{A2.2}
\ee
Thus the distribution of the parameters $\bfp_i$ of the $i$th cloud
can depend on the associated
$\lambda_i$, but not on the $\bfp$'s or $\lambda$'s of any other
clouds.

     Let the
mean density of line centers per unit wavelength interval and in
a volume $d\bfp$ of parameter space be given by $\calN(\lambda,\bfp) d\bfp$.
Generalizing the discussion in \S 5,
we have here allowed the line density function $\calN$ to depend on
wavelength.  For convenience in the following derivations
we shall assume that the mean total density of lines at $\lambda$,
\be
n(\lambda)=\int \calN(\lambda,\bfp)\,d\bfp, \e{A3}
\ee
is finite, temporarily ignoring the possibility (alluded to in the text)
of a power law divergence at small column densities.  We may then take
\be
   P(\bfp_i|\lambda_i)={\calN(\lambda_i,\bfp_i) \over n(\lambda_i)}, \e{A3.1}
\ee
as the
conditional probability
density for the parameters $\bfp_i$ for the $i$th cloud.

    The statistical model defined above is similar to
the well-known {\it Poisson} or {\it shot process} (see, e.g., Parzen 1962).
However, this process has been generalized here to allow
the mean density of centers to vary with $\lambda$ and the function
$T$ to depend on $\lambda$ and $\lambda_i$ in more complicated ways
than simply through their difference.  This generalization is not
strongly needed for the applications of this paper, but will
play a more important role for later papers in this series.

     A very powerful and general way to express the
statistical properties of the function $\tau(\lambda)$ is through its
{\it characteristic functional} (see, e.g., Bartlett 1955).
This is defined for an
arbitrary function $\mu(\lambda)$ by
\be
  \Phi[\mu] \equiv \left\langle
   \exp \left[ i\int \mu(\lambda)\tau(\lambda)\,d\lambda \right]
   \right\rangle,  \e{A4}\ee
where the angle brackets denote averaging over the distribution of
line centers and over the parameter space $\bfp$ for each line.

     The range of integration in equation (\eqref{A4}) has been left
unspecified.  In principle we would like it to be the infinite
range of $\lambda$, but for purposes of the following derivation
it is convenient to choose it temporarily to be some
definite, large (but finite) interval,
which adequately spans the essential nonzero region of the
function $\mu(\lambda)$.
Such a finite region of integration implies
that that the mean number of clouds in the interval
\be      {\Mbar}=\int n(\lambda)\,d\lambda
   =\int d\lambda \int d\bfp \calN(\lambda,\bfp).  \e{A5}\ee
is finite.
The {\it actual} number of lines $M$ in the interval is a random variable with
Poisson distribution function
\be    P_M = {{\Mbar}^M \over M!} e^{-{\Mbar}},  \e{A6}\ee
which can range from $0$ to $\infty$.
For each realization of $M$, the summation in equation (\eqref{A2.1})
is from $i=1$ to $M$.

     We may now evaluate the part of the averaging in equation (\eqref{A4})
due to the Poisson distribution of line centers.   We note that the
probability that a line occurs in a differential interval $d\lambda_i$
about $\lambda_i$ is simply $d\lambda_i\,n(\lambda_i)/{\Mbar}$, and that the
probability distribution for each line is independent of all
other lines (this is the Poisson assumption).  Then
\begin{equation}
  \Phi[\mu] =  \sum_{M=0}^{\infty} {{\Mbar}^M \over M!} e^{-{\Mbar}}
\left\langle \int d\lambda_1\,{n(\lambda_1)\over \Mbar} \cdots
  \int d\lambda_M\, {n(\lambda_M) \over \Mbar}
      \exp\left[i\sum_{i=1}^M \int \mu(\lambda)
     T(\lambda|\lambda_i,\bfp_i)\,d\lambda \right] \right\rangle, \e{A7}
\end{equation}
where now the angular brackets refer only to the average over all the
parameters $\bfp_1$, $\bfp_2$, $\ldots$.  With equations (\eqref{A2.2})
and (\eqref{A3.1}), we have,
\begin{eqnarray}
 \Phi[\mu] &=&  e^{-{\Mbar}}\sum_{M=0}^{\infty} {{\Mbar}^M \over M!}
          \prod_{i=1}^M \left\langle
          \int d\lambda_i\,{n(\lambda_i)\over \Mbar}
          \exp\left[i \int \mu(\lambda)
  T(\lambda|\lambda_i,\bfp_i)\,d\lambda \right] \right\rangle_i \nonumber\\
&=& e^{-{\Mbar}} \sum_{M=0}^{\infty} {1 \over M!}
\prod_{i=1}^M  \int d\lambda_i\,n(\lambda_i)\int d\bfp_i
         {\calN(\lambda_i,\bfp_i) \over n(\lambda_i)}
          \exp\left[i \int \mu(\lambda)
  T(\lambda|\lambda_i,\bfp_i)\,d\lambda \right]  \e{A8}
\end{eqnarray}
We note that all factors in the product are identical,
since the dummy variable of
integration is irrelevant, so the product reduces to a simple power
of one factor, which we may choose to be the one for $i=1$,
\be
  \Phi[\mu]= e^{-{\Mbar}} \sum_{M=0}^{\infty} {1 \over M!}
     \left\lbrace \int d\lambda_1 \int d\bfp_1  \calN(\lambda_1,\bfp_1)
          \exp\left[i \int \mu(\lambda)
  T(\lambda|\lambda_1,\bfp_1)\,d\lambda \right] \right\rbrace^M. \e{A8.1}
\ee
Then using equation (\eqref{A5}) and the power series for the exponential,
along with some changes in dummy variables of integration, we find
\begin{eqnarray}
     \Phi[\mu] &=& \left\langle
   \exp \left[ i\int \mu(\lambda')\tau(\lambda')\,d\lambda' \right]
   \right\rangle \nonumber\\
     &=& \exp\left\lbrace -\int d\lambda
  \int d\bfp\,\calN(\lambda,\bfp)  \left( 1- \exp\left[i\int \mu(\lambda')
         T(\lambda'|\lambda,\bfp)\,d\lambda'
        \right] \right)  \right\rbrace.  \e{A11}
\end{eqnarray}
Recall that finite limits of integration were introduced to make the
quantity $\Mbar$ finite.  However,
assuming a sufficiently localized $\mu(\lambda)$, such divergent
quantities no longer appear in equation (\eqref{A11}), so the integrals in
this formula can be considered to have arbitrary limits.
Similarly, we note that the quantity $n(\lambda)$ no
longer appears, so in many cases this equation can have meaning
even when the integral in equation (\eqref{A3}) diverges.

     Equation (\eqref{A11}) is the major result describing the
statistical properties of the underlying Poisson process.
We can use it to derive a variety of important statistical
results.  For example, if the two sides of the equation are expanded
to first order in $\mu(\lambda')$ and the corresponding terms
equated, one obtains the result for the mean of $\tau(\lambda')$,
\be
 \left\langle \tau(\lambda') \right\rangle
  = \int d\lambda \int d\bfp\, \calN(\lambda,\bfp)T(\lambda' |,\lambda,\bfp).
 \e{A12}
\ee
Equation (\eqref{A11}) can also be expanded to second order.  In doing
so, one must be careful to introduce a new dummy variable of integration
$\lambda''$ on one factor.  Omitting details, we obtain the correlation
function,
\be
  \left\langle \left[ \tau(\lambda')- \langle\tau(\lambda')\rangle \right]
  \left[ \tau(\lambda'')-  \langle \tau(\lambda'')\rangle \right] \right\rangle
 = \int d\lambda \int d\bfp\,
  \calN(\lambda,\bfp)T(\lambda' |,\lambda,\bfp)T(\lambda'' |,\lambda,\bfp).
   \e{A13}
\ee
This result generalizes the well-known Campbell's theorem
(see, e.g., Parzen 1962) by allowing $\lambda$ dependence in
the statistical quantities.

     For the study of \Lya\ clouds, more important than
averages involving $\tau$ itself are averages
involving the exponential extinction law,
\be
         q(\lambda) = e^{-\tau(\lambda)}.  \e{A14}
\ee
Setting $\mu(\lambda')=i\delta(\lambda'-\lambda)$ in equation (\eqref{A11})
gives immediately,
\be
        {\bar q}(\lambda) = \left\langle q(\lambda) \right\rangle =
\exp \left( -\int d\lambda'\int d\bfp \, \calN(\lambda,\bfp)
 \left[ 1-
     e^{-T(\lambda' | \lambda,\bfp)} \right] \right), \e{A15}
\ee
This formula represents a generalization of equation (\eqref{16}) of the
text.

     Correlation properties of $q(\lambda)$ can be found similarly.
Setting $\mu(\lambda''')=i\delta(\lambda'''-\lambda')
+i\delta(\lambda'''-\lambda'')$ in equation (\eqref{A11}) gives
\be
     \left\langle q(\lambda')q(\lambda'') \right\rangle
   =  \exp \left[ -\int d\lambda''' \int d\bfp \calN(\lambda''',\bfp)
     \left( 1- e^{-T(\lambda' |\lambda''',\bfp)-T(\lambda'' |\lambda''',\bfp)}
        \right) \right].  \e{A16}
\ee
[We note that a special case of equation (\eqref{A16}), for $\lambda'
=\lambda''$, and for a homogeneous model, was given by
M\o ller, P., and Jakobsen (1990)].
Using equation (\eqref{A15}) twice, this can be written
\be
     \left\langle q(\lambda')q(\lambda'') \right\rangle =
    {\bar q}(\lambda') {\bar q} (\lambda'')
       e^{-H(\lambda',\lambda'')},  \e{A17}
\ee
where
\be
  H(\lambda',\lambda'') = \int d\lambda''' \int d\bfp \calN(\lambda''',\bfp)
       \left( 1- e^{-T(\lambda' |\lambda''',\bfp)} \right)
       \left( 1 -e^{-T(\lambda'' |\lambda''',\bfp)}  \right).  \e{A18}
\ee
We may also write
\be
     \left\langle\left[ q(\lambda')-{\bar q}(\lambda')\right]
            \left[ q(\lambda'')-{\bar q}(\lambda'')\right] \right\rangle
     = {\bar q}(\lambda') {\bar q} (\lambda'')
       \left( e^{-H(\lambda',\lambda'')} -1 \right).   \e{A19}
\ee
Note that $H(\lambda',\lambda'')$ is small when the wavelength difference
$|\lambda'-\lambda''|$
is much larger than the atomic line widths, since then
the $T$ functions in equation (\eqref{A18}) do not significantly
overlap.  Thus
equation (\eqref{A19}) shows that the values of $q(\lambda)$ are
essentially uncorrelated for such wavelength differences.

\section{Appendix B}

     The purpose of this Appendix is to derive equations
(\eqref{18}) through (\eqref{23}) of \S 5, applying the general
results of Appendix A.

     Before proceeding, we want to take into account
that the measured extinction
is not $q(\lambda)$, but rather an
average over the instrumental profile,
\be
      Q(\lambda) = \int w(\lambda-\lambda')
                         q(\lambda')\,d\lambda'. \e{B1}
\ee
It is usually sufficiently accurate to take $w$ to be
a function of wavelength differences $\lambda-\lambda'$.
In this Appendix, though not in the main text, it is defined
with normalization
\be
          1=\int w(\lambda-\lambda')\,d\lambda'. \e{B2}
\ee

     Besides $Q(\lambda)$ itself, we are also interested in its
square
\be
       Q^2(\lambda) = \int d\lambda'\, w(\lambda-\lambda')
 \int d\lambda''\, w(\lambda-\lambda'') q(\lambda')q(\lambda''),
    \e{B3}
\ee
which is needed in defining the variance of $Q$.  A useful generalization
of this is the product of $Q$'s at two different points,
\be
       Q(\lambda_1)Q(\lambda_2) = \int d\lambda'\, w(\lambda_1-\lambda')
 \int d\lambda''\, w(\lambda_2-\lambda'')
  q(\lambda')q(\lambda''),
    \e{B4}
\ee
which is needed in defining the correlation properties of $Q$.
(The variables $\lambda_1$ and $\lambda_2$ here are general variables,
and have nothing to do with the $i=1$ and $i=2$ clouds.)

     We shall now assume that the distribution $\calN$
is independent of $\lambda$ (or at least that its scale of
variation in $\lambda$ is much larger than either the atomic line profile or
the instrumental resolution).
We also assume that $T(\lambda|\lambda_i,\bfp_i)=
N_i\alpha(\lambda-\lambda_i,b_i)$,
which depends on $\lambda$ and $\lambda_i$ only through their
difference $\lambda-\lambda_i$, at least locally.

     Substituting this expression for $T$ into equation (\eqref{A15}),
we immediately obtain,
\begin{equation}
\left\langle q(\lambda_0) \right\rangle = \left\langle
   e^{-\tau(\lambda_0)} \right\rangle = \exp\left( -\int W(\bfp)
         \calN (\bfp)\,d\bfp \right),  \e{B5}
\end{equation}
where
\begin{equation}
   W(\bfp)= \int d\lambda
   \left\lbrace 1-\exp\left[-N\alpha(\lambda,b)\right] \right\rbrace.
  \e{B6}
\end{equation}
is the equivalent width of a line with column density $N$ and
velocity parameter $b$.
In deriving equation (\eqref{A13}), we have
shifted the variable of integration, showing that
$W(\bfp)$ is independent of $\lambda_0$ (at least over wavelength
intervals sufficiently small that $\alpha$ can be considered to
be only a function of wavelength differences).
Consequently $\langle q \rangle$
is also independent of $\lambda_0$.  Then from equations (\eqref{B1})
and (\eqref{B2}), it follows that
$\overline{Q}\equiv \langle Q \rangle = \langle q \rangle$.
Thus equations (\eqref{B5}) and (\eqref{B6})
constitute an independent proof of equation (\eqref{16}) of the text.

     We next consider the problem of determining the
statistical average of $Q^2$, given in equation (\eqref{B3}).
Actually, it is almost as easy to determine the average of the
product of $Q$'s in equation (\eqref{B4}), so we shall do this.
First of all, we need the
average of $\exp[-\tau(\lambda')-\tau(\lambda'')]$.  This is
given by equation (\eqref{A19}), which can now be written
\be
     \left\langle\left[ q(\lambda')-\overline{Q}\right]
            \left[ q(\lambda'')-\overline{Q}\right] \right\rangle
     = \overline{Q}^2
       \left( e^{-H(\lambda''-\lambda')} -1 \right).   \e{B7}
\ee
where
\be
  H(\lambda) =  \int d\bfp\, \calN(\bfp) \int d\lambda'''
       \left( 1- e^{-N\alpha(\lambda''',b)} \right)
       \left( 1 -e^{-N\alpha(\lambda'''+\lambda,b)}  \right).  \e{B8}
\ee
Here we have made appropriate changes of variables to take advantage
of the difference dependence of the atomic absoption coefficients.
Using equation (\eqref{B4}), we find that
\be
     \left\langle\left[ Q(\lambda_1)-\overline{Q}\right]
            \left[ Q(\lambda_2)-\overline{Q}\right] \right\rangle
     = \overline{Q}^2  \int d\lambda'\, w(\lambda_1-\lambda')
    \int d\lambda''\, w(\lambda_2-\lambda'+\lambda'')
       \left( e^{-H(\lambda'')} -1 \right).   \e{B9}
\ee

     Equation (\eqref{B9}) can be substantially simplified under
conditions where the individual atomic lines are very much
under-resolved, that is, when the instrumental
function $W$ is much broader than atomic widths.
(This is true for the data of the present work.)  In this case,
the effective range of the $\lambda''$ integration is very
much smaller than the width of the instrumental profile, and
we can ignore the $\lambda''$ dependence in the
function $w(\lambda_1-\lambda'+\lambda'')$, setting it equal
to $w(\lambda_1-\lambda')$ and taking it outside the
$\lambda''$ integration.  This gives
\be
     \left\langle\left[ Q(\lambda_1)-\overline{Q}\right]
            \left[ Q(\lambda_2)-\overline{Q}\right] \right\rangle
     = \overline{Q}^2 \left[  \int d\lambda''\, w(\lambda'')
         w(\lambda_2-\lambda_1+\lambda'') \right]
    \int d\lambda''\, \left( e^{-H(\lambda'')} -1 \right).   \e{B10}
\ee
The quantity in brackets determines the dependence of the
correlation function on the wavelength difference $\lambda_1-\lambda$.
When $\lambda_1=\lambda$, the LHS of this equation is equal to
the variance of $Q$, denoted $\hbox{Var}(Q)$.  Thus
\be
     \hbox{Var}(Q) = \overline{Q}^2 \left[  \int d\lambda''\, w^2(\lambda'')
           \right]
    \int d\lambda''\, \left( e^{-H(\lambda'')} -1 \right).   \e{B10.1}
\ee
Thus equation (\eqref{B10}) can be written in the simple form
\be
     \left\langle\left[ Q(\lambda_1)-\overline{Q}\right]
            \left[ Q(\lambda_2)-\overline{Q}\right] \right\rangle
     =  \hbox{Var}(Q) Y(\lambda_2-\lambda_1) ,  \e{B11}
\ee
where
\be     Y(\lambda) = {\int d\lambda''\, w(\lambda'')
         w(\lambda+\lambda'') \over \int d\lambda''\, w^2(\lambda'')}
    \e{B12}
\ee

     For the data of this paper, the instrumental function is
well represented as the rectangular function,
\be
    w(\lambda)=\cases{\Delta^{-1}, &$\lambda < \Delta/2$,\cr
                               0,           &$\lambda > \Delta/2$.\cr}
\e{B13}
\ee
In this case,
\be
     Y(\lambda)=\left( 1-|\lambda|/\Delta \right)_{+}
\e{B14}
\ee
where $(\dots)_{+}$ implies zero for negative arguments.
This shows that the correlation function has a central value
equal to the variance, and has a triangular shape,
going to zero for lags greater than the instrumental width.

     It remains only to determine the variance $\hbox{Var}(Q)$ itself.
For the rectangular instrumental profile this is given by
\be
     \hbox{Var}(Q)/ \overline{Q}^2 =  {1 \over \Delta}
    \int d\lambda''\, \left( e^{-H(\lambda'')} -1 \right).   \e{B15}
\ee

     In general the integral in equation (\eqref{B15})
must be evaluated numerically, which also involves numerical
evaluation of
the function $H(\lambda)$ defined in equation (\eqref{B8}).
However, considerable insight can be obtained by invesigating
the simple model using the single rectangular line profile,
\be
      \alpha(\lambda-\lambda',b)
         =\cases{\alpha_0, &$|\lambda-\lambda'|<W_0/2$, \cr
                 0,        &$|\lambda-\lambda'|>W_0/2$, \cr}
 \e{B16}
\ee
where $\alpha_0$ is the height of the profile and $W_0$ is its full-width.
The parameter $b$ does not appear here, since it implicitly takes
a single value, which sets the values of the constants
$\alpha_0$ and $W_0$.  The parameter vector $\bfp$ then consists
only of the column density $N$.

     With this choice of profile equation (\eqref{B6}) gives
\be
          W(N)=W_0 \left(1-e^{-N\alpha_0}\right),
  \e{B17}
\ee
and equation (\eqref{B5}) then gives
\be
          \overline{Q} = \exp\left[- W_0 \int dN\, \calN(N)
                               \left(1-e^{-N\alpha_0}\right)\right].
    \e{B18}
\ee

     Having found the mean transmission, we next want to find the
variance (and correlation function). From equation (\eqref{B8})
it follows that,
\be
  H(\lambda) = \kappa \left( 1 - |\lambda|/W_0\right)_{+},
   \e{B19}
\ee
where
\be
      \kappa= W_0 \int dN\, \calN(N) \left( 1- e^{-N\alpha_0} \right)^2
   \e{B20}
\ee
The variance is then found from equation (\eqref{B15}),
\be
        \hbox{Var}(Q)/ \overline{Q}^2 =  {2 \over \Delta}
    \int_0^{W_0} d\lambda\, \left( e^{-\kappa(1-\lambda/W_0)} -1 \right)
    = {2W_0 \over \Delta} \left[ {1\over \kappa}\left(e^{\kappa} -1\right)-1
               \right],
  \e{B21}
\ee

     The simplest special case of these results is when there is
also a single column density $N_0$ for all clouds.  Is this case
$\calN(N)=n\delta(N-N_0)$, so that equations (\eqref{B17}) and (\eqref{B18})
imply that the mean transmission is
\be
  \overline{Q} = \exp\left[-n\overline{W}\right],
    \e{B22}
\ee
where the equivalent width of each line is,
\be
               \overline{W}=W_0 \left(1-e^{-N_0\alpha_0}\right).
    \e{B23}
\ee

     The variance is given by (\eqref{B21}), where now,
\be
  \kappa = n W_0\left( 1- e^{-N\alpha_0} \right)^2
            = {n \overline{W}^2 \over W_0},
   \e{B24}
\ee
which proves equation (\eqref{19}) of the text.

     We can also find results for the case where $\calN$ is the power
law in $N$, $\calN(N) = KN^{-\beta}$.  Then
\be
  \overline{Q} = \exp\left[-n\overline{W}\right],
    \e{B25}
\ee
where
\begin{eqnarray}
     \overline{W} &=& K W_0 \int_0^{\infty} dN\,N^{-\beta}
       \left(1-e^{-N\alpha_0} \right) \nonumber\\
     &=& {K W_0 \alpha_0 \over \beta-1}
       \int_0^{\infty} dN\, N^{1-\beta} e^{-N\alpha_0} \nonumber\\
     &=& K W_0 \alpha_0^{\beta-1} \Gamma(2-\beta)/(\beta-1)
   \e{B26}
\end{eqnarray}
An integration by parts has been used, along with the definition
of the $\Gamma$ function.

     Similarly, we now find from equation (\eqref{B20}),
\begin{eqnarray}
   \kappa &=& nK W_0 \int_0^{\infty} dN\,N^{-\beta}
       \left(1-e^{-N\alpha_0} \right)^2 \nonumber\\
          &=& {2nK W_0 \alpha_0 \over \beta-1} \int_0^{\infty}
       dN\,N^{1-\beta} \left( e^{-N\alpha_0}-e^{-2N\alpha_0} \right)\nonumber\\
     &=& (2-2^{\beta-1})n K W_0 \alpha_0^{\beta-1}
              \Gamma(2-\beta)/(\beta-1)
  \e{B27}
\end{eqnarray}
Using equation (\eqref{B26}) we now find
\be
\kappa = (2-2^{1-\beta})n\overline{W}=(2-2^{\beta-1})\ln{1\over \overline{Q}},
   \e{B28}
\ee
which proves equation (\eqref{23}) of the text.

\clearpage
\begin{figure}
\caption{The observed spectra of 29 SSG quasars are here superposed
after shifting each to its emission rest frame and scaling each to
a common magnitude at 1450 \AA.  Despite indisputable differences in
the individual quasars' continuum slopes and emission features, there
is considerable similarity in the spectra.  The principal interest of
this paper is in the statistical analysis of the \Lya\ forest shortward
of 1200 \AA.}
\end{figure}

\begin{figure}
\caption{Fitted line profiles and derived continuum models for each
of the 29 SSG quasars analyzed.  Fitting is done by a linear model,
which can give artifacts in line wings, but is otherwise more robust
than a general nonlinear fit (see text).  The purpose of these fits is
to obtain continuum models that can be extrapolated shortward of 1200 \AA.}
\end{figure}

\begin{figure}
\caption{Extrapolations of the continuum models shown in Figure 2 to
emitted wavelengths between 930 \AA\ and 1200 \AA.  Each point shown,
taken as a fraction of the extrapolated continuum above it, is an
(approximately independent) measurement of \Lya\ absorption at a
calculable redshift.  The ensemble of all the points in this Figure
(excluding a small range of emitted wavelengths around Lyman $\beta$)
is the data set that is analysed in the rest of this paper.}
\end{figure}

\begin{figure}
\caption{Left, the subset of points from Figure 3 with emitted
wavelength between 1050 \AA\ and 1170 \AA\ (each normalized to its
extrapolated continuum level) is plotted as a function of emitted
wavelength.  Since the 29 SSG quasars vary substantially in redshift,
the observed transmission varies widely, with no significant trend.
Right, the same data is plotted as a function of observed wavelength
or, equivalently, absorption redshift for \Lya\ (top scale). Here the
trend with redshift is clear.  The solid line fits a power law model
with mean optical depth varying as $(1+z)^{\gamma+1}$, with
$\gamma=1.46$.  The shaded band approximates the range of statistical
uncertainty in the fit, as determined by the bootstrap method of
resampling the 29 quasars (see text).}
\end{figure}

\begin{figure}
\caption{The individual transmission measurements are shown as
a function of emitted wavelength, after correcting each point for
\Lya\ absorption at its own absorption redshift.  The shaded band
is a moving average of the points.  One sees the Ly-$\beta$,
Ly-$\gamma$, Ly-$\delta$, and possibly Ly-$\epsilon$ decrements.
These are jointly fitted, after shifting each point in the proper
emission region to its proper absorption redshift, and equivalent
width ratios for the corresponding Lyman lines, along with an error
covariance matrix (again obtained by the bootstrap method), are
obtained.  See text for details.  These mean equivalent width ratios
place significant constraints on physical conditions in the clouds.}
\end{figure}

\end{document}